\documentclass[aps,pre,twocolumn,superscriptaddress,showpacs]{revtex4}
\usepackage{epsfig}
\usepackage{amsmath} 

\begin{document}

[Phys. Rev. E {\bf 65}, 065102(R) (2002)]

\title{Topology of the conceptual network of language}

\author{Adilson E. Motter}
\affiliation{Department of Mathematics, Center for Systems Science and Engineering Research,
Arizona State University, Tempe, Arizona 85287}

\author{Alessandro P. S. de Moura}
\affiliation{Department of Mathematics, Center for Systems Science and Engineering Research,
Arizona State University, Tempe, Arizona 85287}
\affiliation{Instituto de F\'{\i}sica, Universidade de S\~{a}o Paulo,
Caixa Postal 66318, 05315-970 S\~{a}o Paulo, Brazil}

\author{Ying-Cheng Lai}
\affiliation{Department of Mathematics, Center for Systems Science and Engineering Research,
Arizona State University, Tempe, Arizona 85287}
\affiliation{Departments of Electrical Engineering and Physics,
Arizona State University, Tempe, Arizona 85287}

\author{Partha Dasgupta}
\affiliation{Department of Computer Science and Engineering,
Arizona State University, Tempe, Arizona 85287}

\date{\today}
 
\begin{abstract}
We define two words in a language to be connected if they express
similar concepts. The network of connections among the many thousands
of words that make up a language is important not only
for the study of the structure and evolution of languages, but also
for cognitive science. We study this issue quantitatively, by mapping out
the conceptual network of the English language, with the connections being
defined by the entries in a Thesaurus dictionary. We find that this network
presents a {\it small-world} structure, with an amazingly small average shortest path,
and appears to exhibit an asymptotic scale-free feature with algebraic 
connectivity distribution.
\end{abstract}
\maketitle

Any language is composed of many thousands of words linked together in an
apparently fairly sophisticated way. A language can thus be regarded as a 
network, in the following sense: (1) the words correspond to nodes of the
network, and (2) a link exists between two words if they express similar
concepts. Clearly, the underlying network of a language is necessarily sparse 
in the sense that the average number of links per node is typically much
smaller than the total number of nodes.  
Identifying and understanding the common network topology of languages  
is of great importance, not only for the study of languages themselves, but
also for cognitive science where one of the most fundamental issues concerns
associative memory, which is intimately related to the network topology.

Recently, there has been a tremendous amount of interest in the study of
large, sparse, and complex networks since the seminal papers by Watts
and Strogatz \cite{WS:1998} on the small-world characteristic and by 
Barab\'{a}si and Albert on scale-free features \cite{BA:1999}. The small-world
concept is {\it static} in the sense that it describes the topological property
of the network at a given time. Two statistical quantities characterizing
a static networks are clustering $C$ and shortest path $L$ , where the former 
is the probability that any two nodes are connected to each other, given that they 
are both connected to a common node, and the latter measures the minimal number
of links connecting two nodes in the network. Regular networks have high
clusterings and small average shortest paths, with random networks at the opposite
of the spectrum which have small shortest paths and low clusterings 
\cite{Bollobas:book}. Small-world networks fall somewhere in between these
two extremes. In particular, a network is small world if its clustering coefficient
is almost as high as that of a regular network but its average shortest path is almost as
small as that of a random network with the same parameters. Watts and Strogatz 
demonstrated that a small-world network can be easily constructed by adding to a
regular network a few additional random links connecting otherwise distant nodes.
The scale-free property, on the other hand, is defined by an algebraic behavior
in the probability distribution $P(k)$ of $k$, the number of links at a node
in the network. This property is {\it dynamic} because it is the 
consequence of the natural evolution of the network. The ground-breaking work
by Barab\'{a}si and Albert \cite{BA:1999} demonstrates that the algebraic 
distribution in the connectivity of scale-free network is caused by two 
basic factors in the temporal evolution of the network: growth and preferential 
attachment, where the former means that the number of nodes in the network
keeps increasing and the latter stipulates that the probability for a new node
to be connected to an existing node is proportional to the number of 
links that this node already has. The scale-free property appears
to be universal for many networks and most of the scale-free networks are also small world.
As of today, the small-world and scale-free features have been discovered in many
networks in nature, and there has also been a large number of theoretical models
proposed to explain these features \cite{Watts:book,AB:2002}.

In this paper, we study the network structure of language \cite{Related_works}. 
We present results for the English language, but they are expected to hold
for any other languages because 
the fundamental role of the language, $i.e.$, to communicate ideas, is shared
by all the languages. We construct a conceptual network from the entries in a 
Thesaurus dictionary and consider two words connected if they express similar
concepts. The network is clearly evolving and sparse. We argue that this network
exhibits the small-world property as a result of natural optimization and,
interestingly, the network 
is asymptotically scale-free due to its dynamic character. We believe and shall
argue that these findings are important not only for linguistics, but also for
cognitive science.

\begin{figure}
\begin{center}
\epsfig{figure=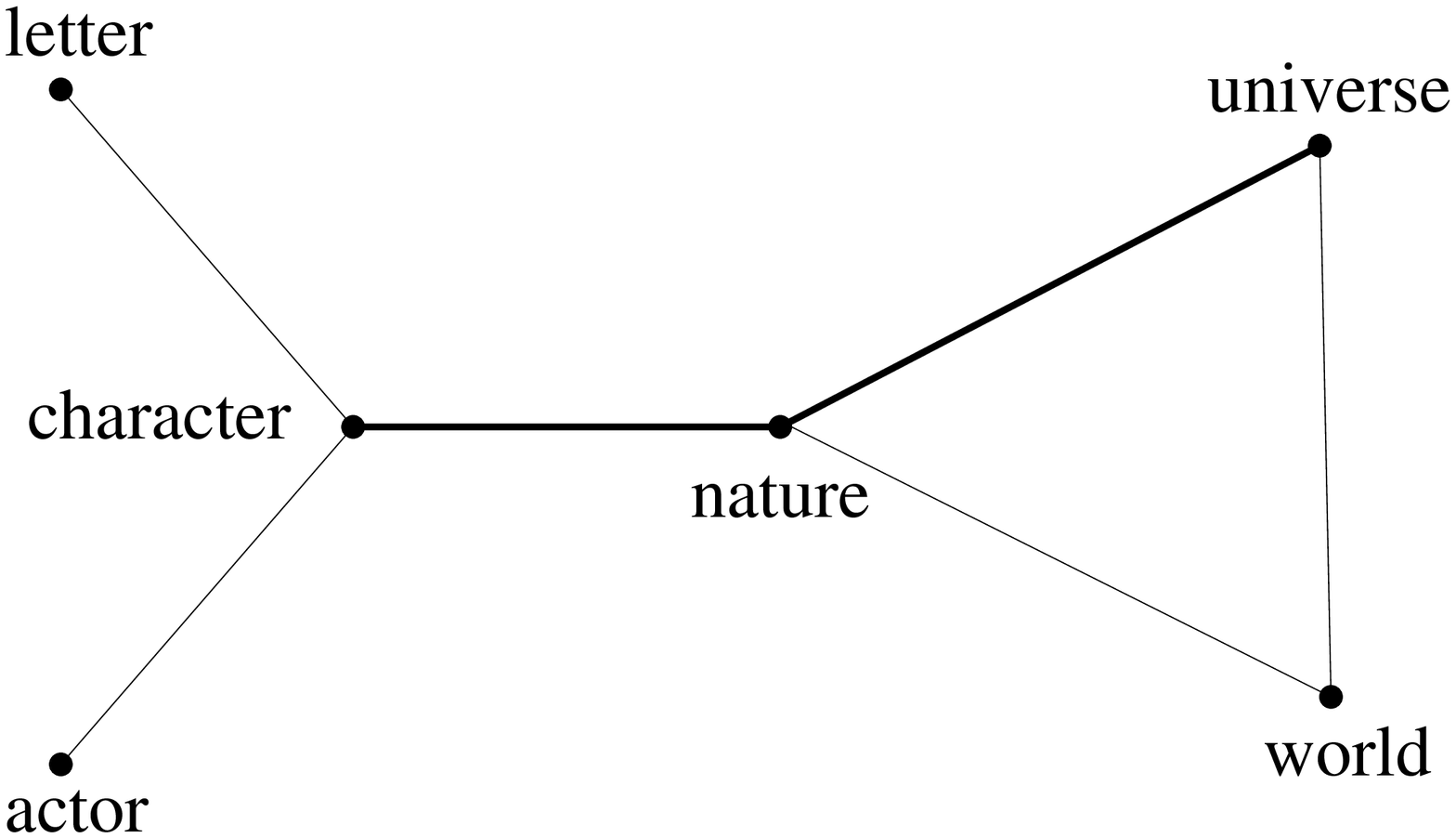,width=\linewidth}
\caption{Illustration of the connections in the conceptual
network for a few words. The thick line is the shortcut between the
words ``universe'' and ``character'', which are connected by
``nature''.}
\label{fig:network}
\end{center}
\end{figure}

A Thesaurus dictionary gives for every entry a list of words that are
conceptually similar to the entry word. For example, the list for the
word ``nature'' includes ``universe'', ``world'', and ``character''. We
define a network from this in a natural way, where each word is a
node, and two nodes are connected if one of the corresponding words is
listed in the entry of the other one. To build this network, we use an
online English Thesaurus dictionary that is freely available \cite{thesaurus},
which has over 30,000 entries, and lists on average over 100 words per
entry. The words that have an entry in the dictionary are called {\em
root words}. Not all words in the list of a given root word are
themselves root words. In the construction of the network, only words
that are root words are considered, and the others are dropped. The
resulting network has an average of about 60 connections per
node. This number is much less than the total number of nodes, and
thus we are dealing with a {\em sparse} network, where each node is
connected to only a small fraction of the network. This is a necessary
condition for the notion of small world to make sense. The construction 
of the network is depicted in Fig. \ref{fig:network}.

We first present results concerning the small-world property of the
network. We expect the network to be  
highly clustered, because there are many sets of related words that
are highly interconnected. For example, ``nature'' is connected to
``universe'', and is also connected to ``world'', and ``world'' and
``universe'' are connected. The numerical calculation of $C$ yields 0.53,
which is compared in Table I with the corresponding value for a random
network with the same parameters, in which the clustering approaches zero,
since the probability that two nodes are connected is
independent on whether they are connected to a common node or not. We
see that in fact $C$ is more than 250 times larger than the random
network value computed from the relation $C=\bar{k}/(N-1)$ \cite{Watts:book}.
On the other hand, because each 
word is linked to only 60 others (on average), compared to over 30,000
in total, and since only words expressing similar concepts are linked,
one might be tempted to conclude that $L$ should be large, and that
one might need to cross hundreds or even thousands of links to go from
one word to another with a very different meaning. However, a
calculation of $L$ yields the amazingly low number of 3.2, which is
very close to the value of about 2.5 of the corresponding random
network estimated from the relation
$L\approx \ln N/\ln \bar{k}$ \cite{Watts:book}, as shown in Table I.
This means that one only needs 3 steps on
average to connect any two words in the 30,000-words dictionary. 

The reason why the average shortest path for the conceptual language
network is so low is related to the existence of words that 
correspond to two or more very different concepts. For example,
``nature'' is connected to ``universe'', but it is also connected to
``character''. Thus, two words with such distinct meanings such as
``universe'' and ``character'' are separated by only 2 links in the
network (c.f. Fig. 1). The word ``nature'' is thus a shortcut that
connects regions of the network that would otherwise be separated by
many links.  The presence of such shortcuts is what makes $L$
small. In fact, less than 1 percent of the words require more than 4
steps to be reached from any given word, on average, as shown in Table II.
For example, one can reach any other word starting from ``nature'' 
with 5 steps or
less.

\begin{table}
\caption{Results for the conceptual network defined by the Thesaurus
dictionary, and a comparison with a corresponding random network with
the same parameters. $N$ is the total number of nodes (root words),
$\bar{k}$ is the average number of links per node, $C$ is the clustering
coefficient, and $L$ is the average shortest path.}
\begin{center}
\begin{tabular}{|l|l|l|l|l|}
\hline
&  
$N$  &  $\bar{k}$  &  
$C$  &  $L$  \\
\hline 
Actual configuration &
30,244  &  59.9  &
0.53   &  3.16 \\
\hline
Random  configuration &
30,244  &  59.9  &
0.002  &  2.5  \\
\hline
\end{tabular}
\end{center}
\end{table}

Our first result is thus that the conceptual network is highly clustered and
at the same time has a very small length, $i.e.$, it is a {\em
small-world network}. Since the length $L$ in
small-world networks grows only logarithmically with the number of
nodes \cite{WS:1998}, even if we included more words in the dictionary (and
consequently more nodes), $L$ would not change by much, and our
conclusions still hold. Another important point is that even though we
used the dictionary of a particular language (English), since the
Thesaurus associates words based on their concepts, we expect similar
results to hold for other languages as well. In fact, in any language the
network will be highly clustered, and any language has words that
function as shortcuts, guaranteeing that $L$ is very small, even though
the particular words that act as shortcuts may be different for
different languages.

\begin{table}
\caption{Average number $N_n$ of nodes at a shortest path $L=n$ from a given node
in the conceptual network. $\rho\equiv N_n/N$ is the
fraction of nodes corresponding to $N_n$.}
\begin{center}
\begin{tabular}{|c|c|c|}
\hline   
$\;\;\; n\;\;\;$  & $N_n$ & $\rho$ \\
\hline 
 1  &  59.9 & 0.002 \\
\hline
 2  &  2,961 & 0.098 \\
\hline
 3  &  19,762 & 0.653 \\
\hline 
 4  &  7,205 & 0.238 \\
\hline
 5  &  222 & 0.007 \\
\hline
 6  &  28.5 & 0.001 \\
 \hline
 7  &  4.7 & $\sim 10^{-4}$ \\
\hline
 8  &  0.06 & $\sim  10^{-6}$ \\
\hline
\end{tabular}
\end{center}
\end{table}

\begin{figure}
\begin{center}
\epsfig{figure=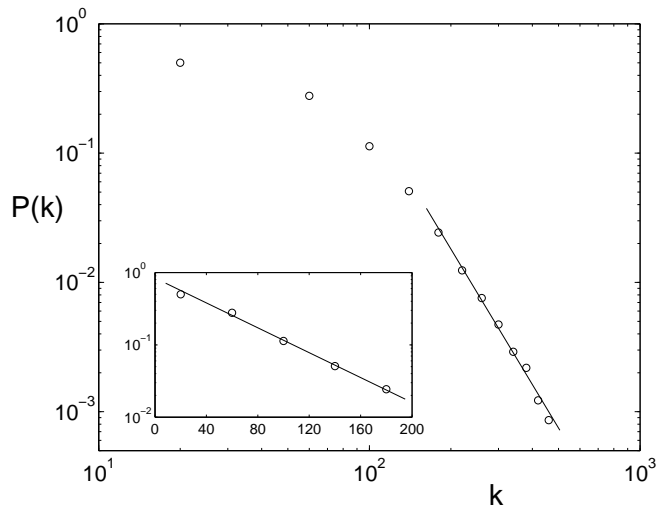,width=\linewidth}
\caption{Algebraic scaling behavior of $P(k)$ for the conceptual network of the English language.
The inset shows the initially exponential decay of $P(k)$.}
\label{fig:scaling}
\end{center}
\end{figure}

Next we consider the dynamical feature of the conceptual network.
The language is an evolving system, where new words are continually
created and added to the network. The conceptual network of language
can thus be regarded as a growing network. 
But, how are the new nodes attached in the conceptual network?
The answer is encoded in the probability distribution $P(k)$ of the 
connectivity. If new nodes are randomly added to the network, $P(k)$
follows an exponential distribution \cite{BAJ:1999}: $P(k)\sim \exp (-\beta k)$.
If new nodes are preferentially added to the network, $e.g.$, if the 
probability $\Pi_i$ for an already existing node $i$ to acquire a link
from the new node is proportional to $k_i$, the number of links that
node $i$ already has, then $P(k)$ exhibits the following algebraic scaling
\cite{BA:1999,BAJ:1999}:
\begin{equation} \label{eq:scaling}
P(k) \sim k^{-\alpha},
\end{equation}
where $\alpha = 3$. The algebraic scaling law (\ref{eq:scaling}) reflects
the fact that there is a self-organizing principle governing the growth of
the network, which has indeed been discovered in many realistic networks
\cite{BA:1999,AB:2002}. For our conceptual network of language, we expect
the distribution $P(k)$ to reflect the intrinsically coherent manner by
which a language is supposed to evolve. However, the rule of a perfect 
preferential attachment $\Pi_i \sim k_i$ appears to be too idealized as
there are also random factors affecting how a new word is added to the 
language. We thus hypothesize that for the conceptual network
of language, a new node is added to the network with both preferential
and random attachments. Specifically, we assume,
\begin{equation} \label{eq:Pii}
\Pi_i \sim (1-p)k_i + p,
\end{equation}
where $p$ and $(1-p)$ are the weights of random and preferential attachments,
respectively. A recent work \cite{LLYD:2002} indicates that the attachment 
rule (\ref{eq:Pii}) leads to the following connectivity distribution:
\begin{equation} \label{eq:crossover}
P(k) \sim (k+\frac{p}{1-p})^{-\gamma}, \;\;\; \gamma=3+\frac{p}{m(1-p)},
\end{equation}
where $m$ is the number of new links added to the network at each time step.
We see that for small $k$, $P(k)$ exhibits an approximately exponential behavior, while
for large $k$, $P(k)$ appears to be algebraic with an exponent greater than 3. We then
expect to observe a {\it crossover} from the exponential to algebraic behavior as
$k$ is increased. This indeed appears to be the case for the conceptual network of language,
as shown in Fig. \ref{fig:scaling}, where the asymptotic algebraic scaling exponent
is about 3.5, which is consistent with the theoretical prediction in 
Eq. (\ref{eq:crossover}). This indicates that our hypothesis of mixed contributions
from preferential and random attachments in the development of the conceptual 
network of language is plausible, and there is indeed a self-organized structure
in the network to certain degree.   

A heuristic justification for our hypothesis (\ref{eq:Pii}) is as follows.
Because of the small-world topology, each node of the conceptual network on average has a 
large fraction of local connections and a small fraction of long range connections.
When a new node is added to the network, it has the same probability of attaching to
any one of the already existing nodes. But, once it attaches a node $j$ it has the
tendency to connect preferentially to the nodes that are already connected to $j$ \cite{Social_network}.
Preferential attachment comes from the second step, since the probability that a node $i$ is in the
neighborhood of node $j$ is proportional to the number of links $k_i$ of node $i$; while
the random component comes from the random choice of the first connection $j$ and the subsequent
long range connections.
The small-world property is consistent with the evolutionary character of the network,
as the growing process tends to keep high clustering and small shortest path.

In comparison with the small-world model originally proposed in Ref. \cite{WS:1998},
a scale-free network presents a highly heterogeneous distribution of links per node.
In spite of this, the evolution of the conceptual network is demonstrated to be robust, 
in that most of the words correspond to nodes connected to few other nodes, and can be
removed without affecting the structure of the network \cite{AJB:2000,LLYD:2002}.  
There are also words that are the most visible ones, but they are unlikely to be suddenly 
lost or undergo an abrupt transformation in the evolution without a self-organized
reconnection of the neighbors \cite{NMLH:2002}.

We conclude with some thoughts on the meaning of our results for
cognitive science. It is well known that human memory is associative,
which means that information is retrieved by connecting similar
concepts, just as in our network above \cite{KL:1999,HSW:2001}. From the standpoint of
retrieval of information in an associative memory, the small-world
property of the network represents a maximization of efficiency: on
the one hand, similar pieces of information are stored together, due
to the high clustering, which makes searching by association possible;
on the other hand, even very different pieces of information are never
separated by more than a few links, or associations, which guarantees a
fast search. We thus speculate that associative memory has arisen
partly because of a maximization of efficiency in the retrieval by
natural selection. This issue may be related to the fact that the neural
network is probably a small-world network as well \cite{SHBNYK:2000,LM:2001},
which is probably necessary for the brain to be able to hold a conceptual network
that is needed for associative memory.

This work was supported by FAPESP and AFOSR CIP (Critical Information Protection)
Program under Grant No. F49620-01-1-0317.


\begin{references}
\bibitem{Social_network}
This model can be better motivated in the context of social networks.
In a friendship network, when a new person is introduced into the group,
he or she has uniform probability to make a first friend.
The probability of making new friends, however, will depend on this
first connection, since a person is more likely to be introduced to
the friends of his or her friends than to any other person.
\end{references}
\end{document}